\documentclass[aps,twocolumn,showpacs,groupedaddress]{revtex4}
\usepackage{graphicx}

\begin{document}

\title{Quantum dynamics in light-harvesting complexes: Beyond the single-exciton limit}
\author{B. Cui, X. Y. Zhang and X. X. Yi}
\affiliation{School of Physics and Optoelectronic Technology,\\
Dalian University of Technology, Dalian 116024 China}

\date{\today}

\begin{abstract}
Primitive photosynthetic cells appear over three billion years prior
to any other more complex life-forms, thus it is reasonable to
assume that Nature has designed a photosynthetic mechanism using
minimal resources but honed to perfection  under the action of
evolution. A number of different quantum  models have been proposed
to understand the high degree of efficient energy transport, most of
them are limited to the scenario of single-exciton. Here we present
a study on the dynamics in light-harvesting complexes beyond the
single exciton limit, and show how this model describes the energy
transfer in the Fenna-Matthew-Olson (FMO)  complex. We find that the
energy transfer efficiency above 90\% under realistic
conditions is achievable.
\end{abstract}

 \maketitle
\section{introduction}
Light-harvesting complexes consist of several chromophores mutually
coupled by dipolar interactions residing within a protein scaffold.
Due to their mutual coupling, light-induced excitations on
individual chromophores (sites) can undergo  transfer from site to
site. Excitation  energy transfer (EET) has been an interesting
subject  for decades, not only for its phenomenal efficiency but
also for its fundamental role in Nature \cite{foster48}. Recent
experiments on the  exciton dynamics in photosynthetic bio-molecules
(for example, the purple bacteria  and the Fenna-Matthew-Olson
complex) have brought a long-standing question again into the
scientific focus that whether nontrivial quantum coherence effects
exist in natural biological systems under physiological conditions
\cite{lee07,adolphs06}. In fact, evidence of quantum coherence has
been found, suggesting  that nontrivial quantum effects may be at
the heart of its remarkable excitation transport efficiency \cite{engel07}.

Inspired  by these experimental results, several studies have
attempted to unravel the precise role of quantum coherence in the
EET of light-harvesting complexes
\cite{mohseni08,plenio08,caruso09,chin10,olaya08,ishizaki09,yang10,hoyer09,sarovar11},
and environmental decoherence and noise have been found to play a
crucial role
\cite{mohseni08,plenio08,caruso09,chin10,caruso10,shabani11}. In
these studies, the system is assumed to be initialized with a single
excitation in site 1. This may not be realized precisely under
experimental or natural operating conditions. Considering that
primitive photosynthetic cells appeared over three billion years
prior to any other more complex life-forms, it is not illogical to
assume that nature has designed a photosynthetic mechanism using
minimal resources to gain  maximal energy  under the action of
evolution. In this perspective, a model which allows a freedom to
control the number of excitations in the complex at any time should
be taken into account.
\begin{figure}
\includegraphics*[width=0.6\columnwidth,
height=0.7\columnwidth]{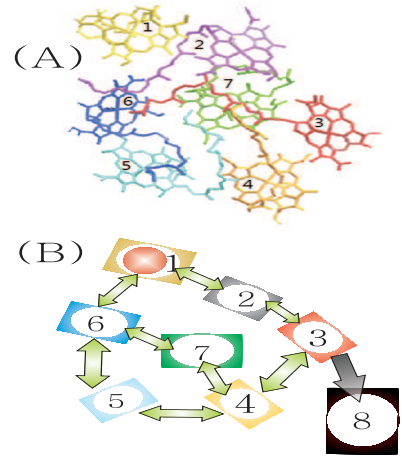}
\caption{(A) The disordered structure of the Fenna-Matthews-Olson
(FMO) complex. The FMO complex acts as an energy transfer channel in
green sulphur bacteria guiding excitations from the lightharvesting
antenna at site 1 to  the reaction center at site 8. This picture is
re-produced from \cite{shabani11}. (B) The model to describe the FMO
complex. Arrows between the cavities represent cavity-cavity
couplings. Only couplings above 15 $\mbox{cm}^{-1}$ are shown. The
reaction center is modeled by the cavity, numbered 8, that is
irreversibly coupled to the site 3.  It is worth bearing in mind that the coupling
between cavity 1 and cavity 6 is  week with respect to the other
couplings shown in the figure.} \label{fig1}
\end{figure}

This paper extends the theoretical formulation presented in a recent
paper \cite{caruso09} to a scenario of multi-exciton  and further
examines issues relevant for realistic light-harvesting complexes.
To this end, we identify the FMO complex with coupled cavities and
introduce two fundamental decoherence mechanisms (i.e., dephasing
and dissipation) into the system. Under the semi-classical approach
and with the quantum theory, we calculate respectively the
excitation  transfer efficiency. The decoherence rates that are
optimal to the ETE are found by numerical simulation of the equation
of motion. With the optimal decoherence rates, the time evolution of
population on each site is presented. We found that the optimal
decoherence rates weakly depend on the initial excitation number on
site 1. The non-local decoherence slightly alter the ETE, indicating
that local decoherence dominates the energy transfer in the FMO
complex.

The paper is organized as follows. In Sec. {\rm II} we introduce the
theoretical model for the FMO dynamics and define the excitation
transfer efficiency (ETE) used here. Then, we analyze the ETE and
the dynamics in the FMO complex with both semi-classical and quantum
theories in Sec.{\rm III}. Finally, we conclude our results in Sec.
{\rm IV}.

\section{Model description}
In photosynthetic antennae, sunlight is absorbed by pigments and the
excitation energy is transferred to the photosynthetic reaction
center. The locations for these processes are physically and
physiologically separated, suggesting that two-level systems are
good to model the pigments with single-exciton but  not enough for
pigments with many excitations. To describe  the light-harvesting
process with many excitations on each site, we model the FMO complex
by a coupled 8-cavity system, see Fig. \ref{fig1}. The effective
dynamics can be described by a Hamiltonian describing  the coherent
exchange of excitations between chromophores or sites,
\begin{equation}
H=\sum_{j=1}^7\omega_j a_j^{\dagger}a_j+\sum_{i,j=1}^7
 g_{ij}(a_i^{\dagger} a_j+ a_ia_j^{\dagger}),\; i\neq j\; ,
\end{equation}
where $a_j^{\dagger}$  ($a_j$) are the creation  (annihilation)
operators for site $j$, $\omega_j$ is the local site excitation
energy, and $g_{ij}$ denotes the hopping rate of an excitation
between the sites $i$ and $j$. In the site basis, we follow
\cite{adolphs061} and employ the Hamiltonian matrix elements (in
units of $\mbox{cm}^{-1}$)
\begin{widetext}
\begin{equation}
        H \!=\!\! \left(\!\!\begin{array}{rrrrrrr}
         \mathbf{215}   & \!\mathbf{-104.1} & 5.1  & -4.3  &   4.7 & \mathbf{-15.1} &  -7.8 \\
        \!\mathbf{-104.1} &  \mathbf{220.0} &\mathbf{ 32.6} & 7.1   &   5.4 &   8.3 &   0.8 \\
           5.1 &  \mathbf{ 32.6 }&  0.0 & \mathbf{-46.8} &   1.0 &  -8.1 &   5.1 \\
          -4.3 &    7.1 &\!\mathbf{-46.8} & \mathbf{125.0} &\! \mathbf{-70.7} &\! -14.7 &
          \mathbf{ -61.5}\\
           4.7 &    5.4 &  1.0 & \!\mathbf{-70.7} & \mathbf{450.0} & \mathbf{ 89.7} &  -2.5 \\
         \mathbf{-15.1} &    8.3 & -8.1 & -14.7 &  \mathbf{89.7} & \mathbf{330.0} & \mathbf{ 32.7} \\
          -7.8 &    0.8 &  5.1 & \mathbf{-61.5} &  -2.5 &  \mathbf{32.7} & \mathbf{280.0}
          \end{array}\!\!
        \right).
        \label{ha}
\end{equation}
\end{widetext}
Here the zero energy has been shifted by 12230 $\mbox{cm}^{-1}$ for
all sites, corresponding to a wavelength of $\sim 800 \mbox{nm}$. We
note that in units of  $\hbar=1$,  1 ps$^{-1}$=5.3 cm$^{-1}$. Then
 by dividing $g_{ij}$ and $\omega_{j}$ by 5.3, all elements of the Hamiltonian  are
 rescaled in units of
ps$^{-1}$. We can find from the Hamiltonian $H$ that in the
Fenna-Matthew-Olson complex (FMO), there are two dominating EET
pathways: $1\rightarrow 2\rightarrow 3$ and $6\rightarrow
(5,7)\rightarrow 4 \rightarrow 3$ (see figure \ref{fig1}). Although
the nearest neighbor terms dominate the site to site coupling,
significant hopping matrix elements exist between more distant
sites. This indicates that coherent transport itself may not explain
why the excitation energy transfer is so efficient.

To obtain high energy transfer efficiency in the EET process,
forward and backward energy transfer rates as well as the
dissipation induced by the {\it hot and wet} surrounding environment
need to satisfy a detailed balance condition. In the weak
dissipation regime, the Lindblad master equation that is able to
reliably describe exciton dissipative dynamics reads,
\begin{eqnarray}
        \frac{d\rho}{dt} &=& -i[H,\rho]
        + {\cal L}_{deph}(\rho) + {\cal L}_{diss}(\rho)\nonumber\\
        && + {\cal L}_{NLdeph}(\rho) + {\cal L}_{NLdiss}(\rho) + {\cal L}_{8}(\rho)\;,
\label{masterE}
\end{eqnarray}
where ${\cal L}_{diss}(\rho)$ describes dissipation terms, and
${\cal L}_{deph}(\rho)$ denotes dephasing terms. ${\cal
L}_{NLdeph}(\rho)$ and ${\cal L}_{NLdiss}(\rho)$ represent non-local
dephasing and dissipation, respectively. Here,
\begin{equation}
{\cal L}_{diss}(\rho)=\sum_{j=1}^7 \Gamma_{j}(2a_j\rho
a_j^{\dagger}-\rho a_j^{\dagger}a_j-a_j^{\dagger}a_j\rho)\;,
\end{equation}
\begin{equation}
{\cal
L}_{deph}(\rho)=\sum_{j=1}^7\gamma_j(2n_j{\rho}n_j-\rho{n_j}n_j-n_jn_j\rho),
\end{equation}
\begin{equation}
{\cal L}_{NLdiss}(\rho)=\sum_{i,j=1}^7 \Gamma_{ij}(2a_i\rho
a_j^{\dagger}-\rho
a_j^{\dagger}a_i-a_j^{\dagger}a_i\rho),i\neq{j},\label{Ndiss}
\end{equation}
and
\begin{equation}
{\cal L}_{NLdeph}(\rho)=\sum_{i,j=1}^7\gamma_{ij}(2n_i{\rho}n_j
-\rho{n_j}n_i-n_jn_i\rho),i\neq{j},\label{Ndeph}
\end{equation}
where $n_j=a_j^{\dagger}a_j$ denotes the exciton number operator for
site $j$.

The local decoherence may come from the couplings of the sites to
individual environments, whereas the non-local decoherence  ${\cal
L}_{NLdeph}(\rho)$ and ${\cal L}_{NLdiss}(\rho)$ can be understood
as a result of the interaction between the site and a common
environment. From a quantum physical perspective, environmental
radiations whose wavelength is larger than the length-scale of the
FMO molecule ($\sim 8$ nm) are reasonably assumed to take the role
of a common environment. Thus the decoherence considered here is
reasonable.

Recent work \cite{adolphs06} suggests that it is the site 3 that
couples to the reaction center. The total transfer of excitations
into the reaction center is measured by the population in the
center, numbered 8, which is populated by an irreversible decay
process with  rate $\Gamma_{8}$ from the site 3. We
phenomenologically model this irreversible process   by the Lindblad
operator,
\begin{equation}
{\cal L}_{8}(\rho)=\Gamma_{8}(2a_8^{\dagger} a_3\rho
a_3^{\dagger}a_8-\rho a_3^{\dagger}a_3a_8a_8^{\dagger}-
a_3^{\dagger}a_3a_8a_8^{\dagger}\rho)\;.
\end{equation}
To match the observation in experiments, we assume there are $N_0$
excitations initially in site 1. The model is completed by
introducing a quantity by which we measure the energy transfer
efficiency.  The rescaled population in the reaction center given by
\begin{equation}
p_8=\frac{n_8(T)}{N_0}=\frac{\mbox{Tr}(a_8^{\dagger}a_8\rho(T))}{N_0},
\label{effi}
\end{equation}
at a specific time $T$ is good for this purpose.

\begin{figure}
\begin{minipage}[t]{0.48\linewidth}
\includegraphics*[width=1\columnwidth,
height=0.7\columnwidth]{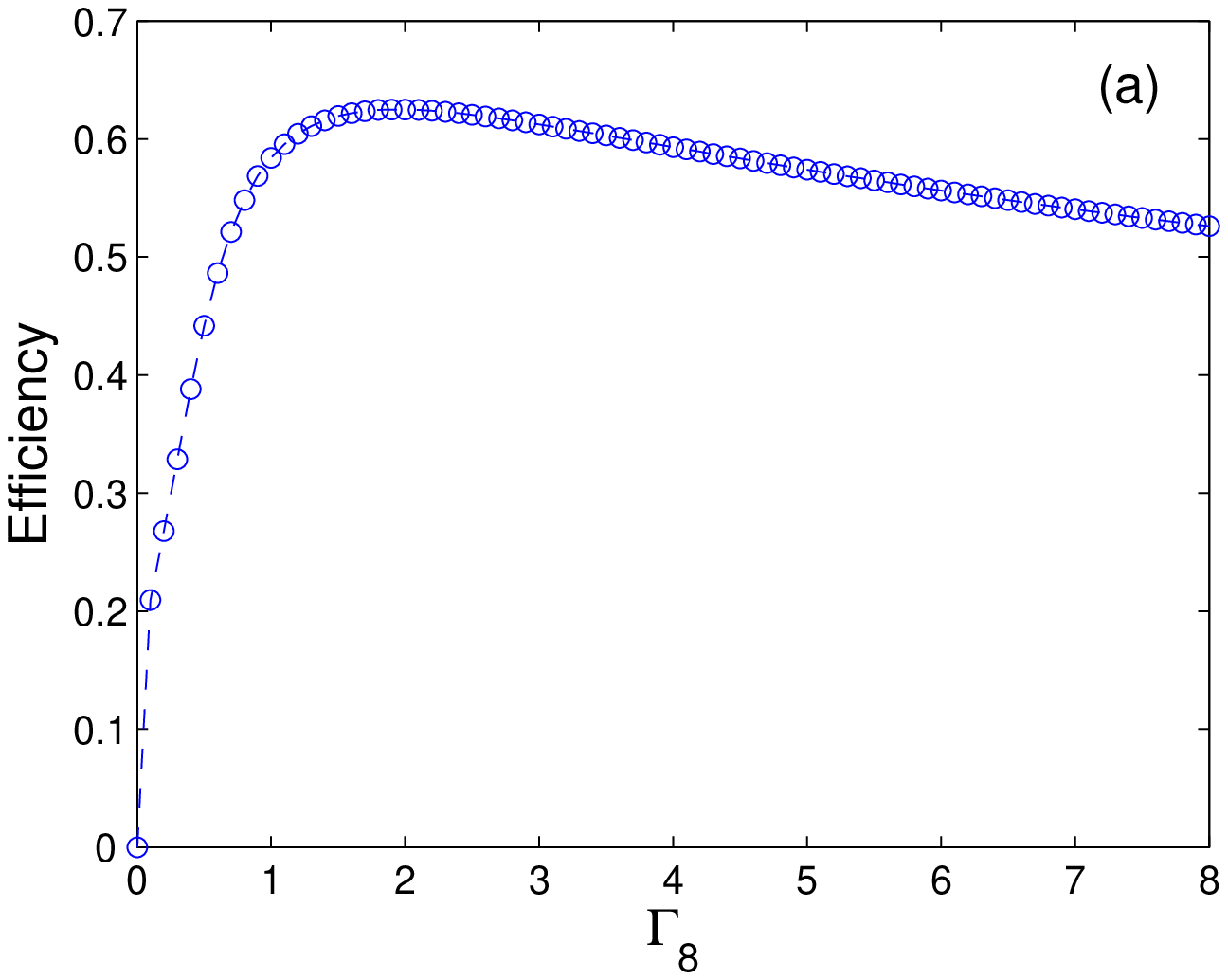}
\end{minipage}
\begin{minipage}[t]{0.48\linewidth}
\includegraphics*[width=1\columnwidth,
height=0.7\columnwidth]{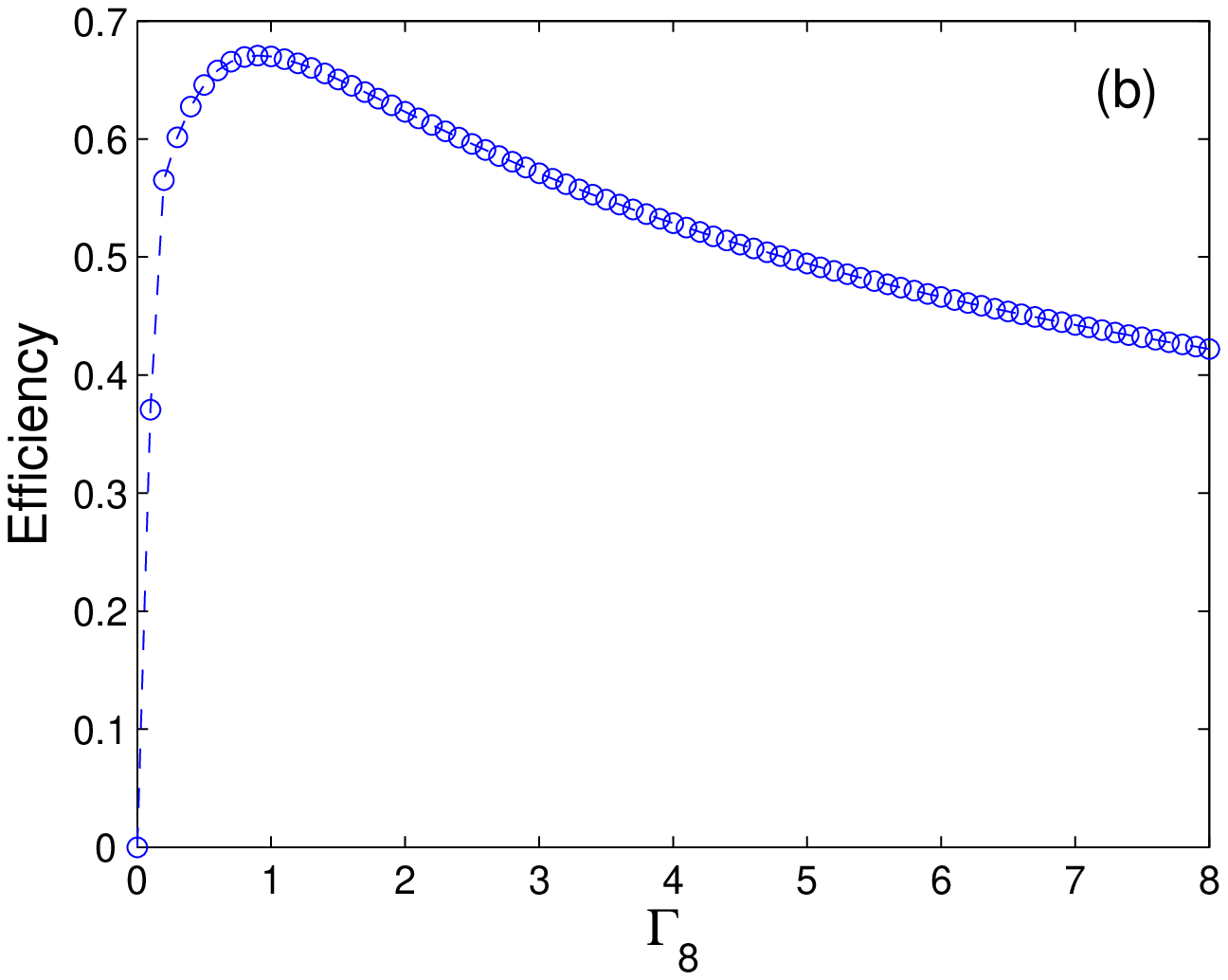}
\end{minipage}
\caption{Energy transfer efficiency versus $\Gamma_8$ without
decoherence ($\Gamma_j=\gamma_j=\Gamma_{ij}=\gamma_{ij}=0,$
$i,j$=1,2,...,7). (a) For semi-classical approach  and, (b) for
quantum theory. $N_0=100$ is taken for this plot. }\label{fig2}
\end{figure}

\begin{figure}
\begin{minipage}[t]{0.48\linewidth}
\includegraphics*[width=1\columnwidth,
height=0.8\columnwidth]{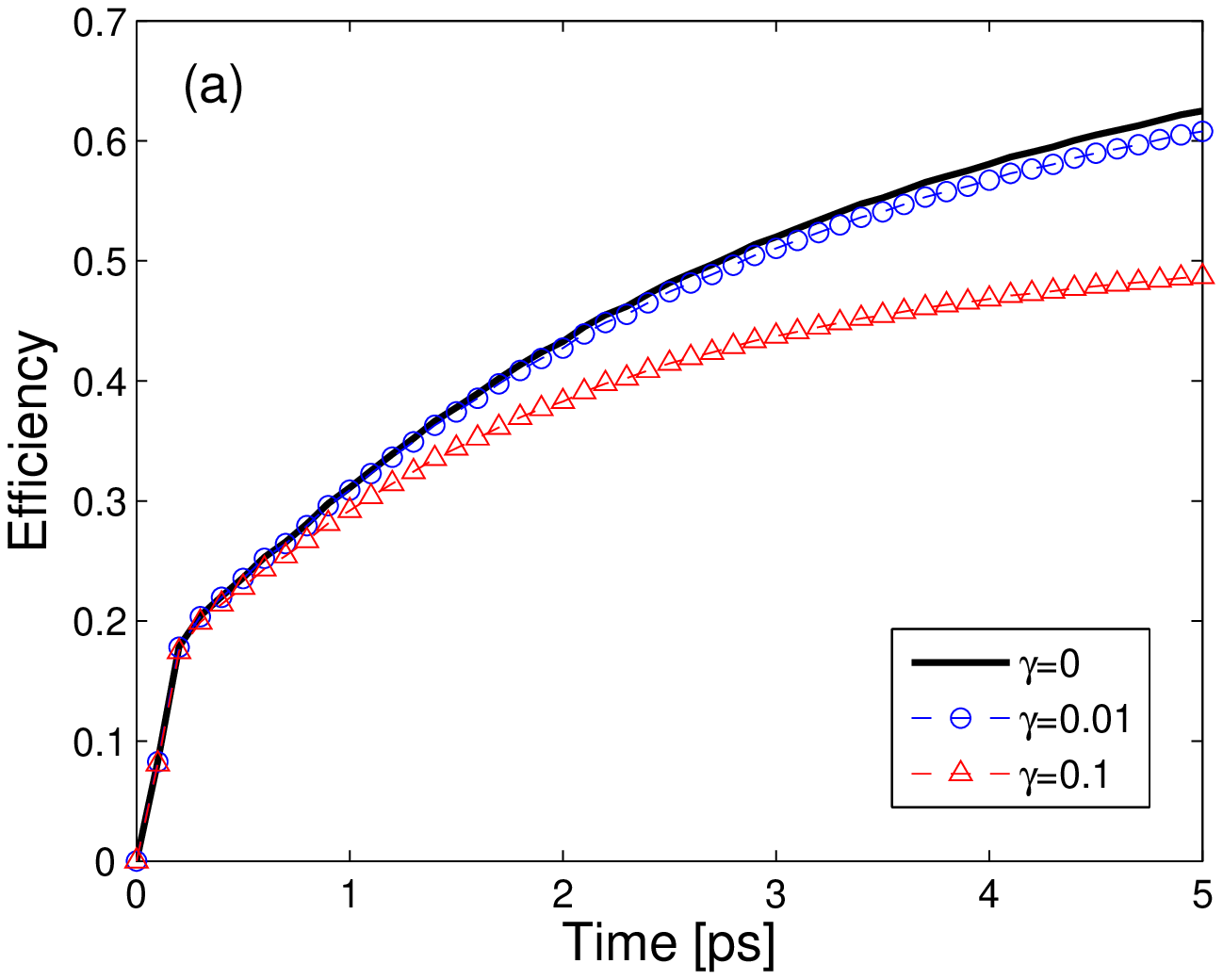}
\end{minipage}
\begin{minipage}[t]{0.48\linewidth}
\includegraphics*[width=1\columnwidth,
height=0.8\columnwidth]{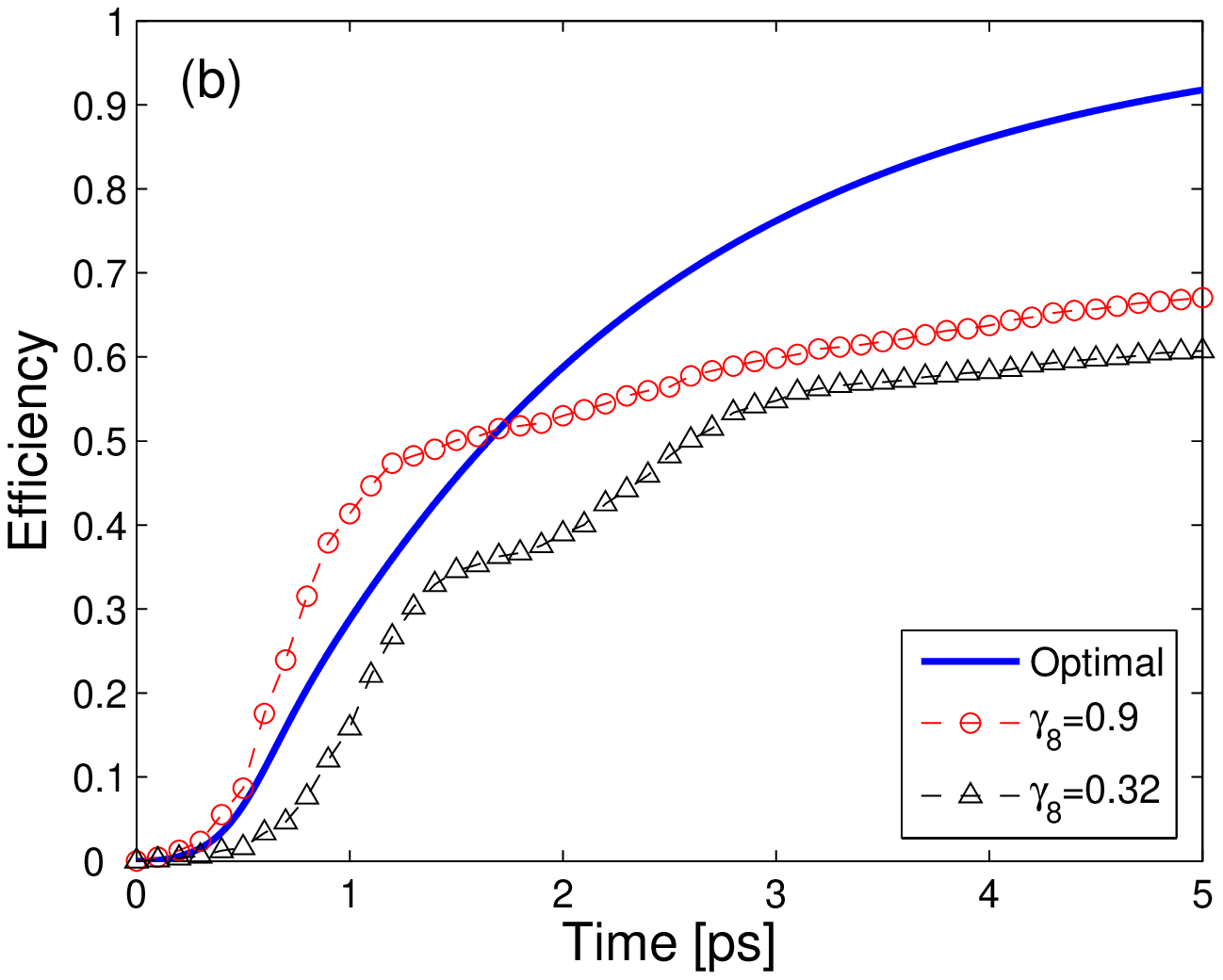}
\end{minipage}
\caption{(color online) Efficiency as a function of time, only local
decoherence is considered. Figure (a)  is  the results with
semi-classical approach, where black solid line is for the null
decoherence case ($\Gamma_j=\gamma_j=0$), and  blue circle and red
triangle are for $\Gamma_j=0,$ $\gamma_j=0.01$ and $\Gamma_j=0,$
$\gamma_j=0.1,$ respectively, $j=1,2,...,7$. Other parameters
chosen are $\Gamma_8=1.94$ and $N_0=100$. (b) is the efficiency from
quantum theory. Blue solid line stands for  optimal efficiency with
rates ($\gamma_1, \gamma_2, \gamma_3, \gamma_4, \gamma_5, \gamma_6,
\gamma_7$) =(0.74, 24, 0, 5.2, 50.6, 0, 15), $\Gamma_8=0.32$ and the
other $\Gamma_j=0.0005$. The efficiency reaches its maximum
91.77$\%$ that is optimized  at 5 ps. Red circle and black triangle
lines represent the null decoherence case with $\Gamma_8=0.9$ (the
corresponding efficiency is 67.06$\%$) and $\Gamma_8=0.32$
(corresponding to  efficiency 60.73$\%$), respectively. }
\label{fig3}
\end{figure}
Note that the reaction center is directly connected only to site 3.
The decoherence rate $\Gamma_8$ plays an essential role in the
excitation  transfer. In the next section, we will use the
decoherence rates to optimize the transfer efficiency defined in
Eq.(\ref{effi}). Several cases are considered, in each case
$\Gamma_8$ can not be zero, because null $\Gamma_8$ leads to zero
transfer efficiency.

\section{energy transfer efficiency}
Having this model, we  now explore the energy transport in the FMO
complex. Two approaches   are considered. In the semi-classical
approach, we will use the approximation $\langle a_j^{\dagger}
a_j\rangle =|\alpha_j|^2$ with $\alpha_j=\mbox{Tr}(\rho a_j).$
Whereas in the quantum regime, we approximate $\langle a_8^{\dagger}
a_8 a_3^{\dagger} a_3\rangle$ as $\langle a_8^{\dagger} a_8\rangle
\langle a_3^{\dagger} a_3\rangle.$  By these approximations, we can
derive the equation of motion for the system and  calculate  the
energy transfer efficiency. We first focus on the case where only
local decoherence exists, namely $\Gamma_{ij}=0$ and $
\gamma_{ij}=0$, then move to the case  with non-local decoherence.
The energy transfer time is taken to  $T$ = 5 ps$^{-1}$, which is
the relevant time scale in recent experiment \cite{adolphs06}. Our
results suggest that it is the careful interplay of quantum
mechanical features and the unavoidable couplings to environment
that will leads to the optimal system performance. In particular,
the local decoherence dominates and actually helps the excitation
transfer, as we will show below.

\subsection{Semi-classical approach}
 Define $\alpha_j=\mbox{Tr}(\rho a_j)$ and
$n_8=\mbox{Tr}(a_8^{\dagger}a_8\rho)$, the master equation Eq.
(\ref{masterE}) yields,
\begin{widetext}
\begin{eqnarray}
\dot{\alpha}_{j}&=&-i({\omega_j}{\alpha_j}+2\sum_{k\neq{j}}{g_{jk}\alpha_k})-
\Gamma_j{\alpha_j}-\gamma_j{\alpha_j}-\sum_{k\neq j}\Gamma_{jk}\alpha_k,j\neq3,8,\nonumber\\
\dot{\alpha}_{3}&=&-i({\omega_3}{\alpha_3}+2\sum_{k\neq{3}}{g_{3k}\alpha_k})-
\Gamma_3{\alpha_3}-\gamma_3{\alpha_3}
-\sum_{k\neq3}\Gamma_{3k}\alpha_k-\Gamma_8(n_8+1)\alpha_3,\nonumber\\
\dot{n}_8&=&2\Gamma_8|\alpha_3|^2(n_8+1),
\label{eofm}
\end{eqnarray}
\end{widetext}
where the dot represents time derivative, and the time arguments
have been omitted to shorten the representation. To obtain Eq.
(\ref{eofm}), the approximation $\langle a_j^{\dagger} a_j\rangle
=|\alpha_j|^2$ in the last equation of Eq. (\ref{eofm}) has been
made. These equations compose a closed set of equations governing
the dynamics of the FMO complexes.

As shown in \cite{caruso09}, a completely coherent dynamics is often
not most ideal for the excitation transfer  from the chromophores to
the reaction center,  and hence the coherence solely  can not
explain the very high exciton transfer efficiency observed in
experiments. Our model confirms this result. In fact, with
$\Gamma_j=\gamma_{j}=\Gamma_{ij}=\gamma_{ij}=0$ ($i,j=1,2,...,7$),
the transfer efficiency arrives at its maximum $0.625$ with
$\Gamma_8=1.94$ ps$^{-1}$ (see Figs. \ref{fig2}(a) and
\ref{fig3}(a)).

Further numerical simulations show that under the semi-classical
approximation, dissipation  and dephasing  play similar roles  in
the  dynamics of FMO, see Eq. (\ref{eofm}). Numerical simulations
show that neither dissipation nor  dephasing can  improve the energy
transfer efficiency under the semi-classical approximation (see Fig.
\ref{fig3}(a)). This is due to the fact that dephasing and
dissipation play a same role in the dynamics under the
semi-classical approach, see Eq. (\ref{eofm}). Remind that the local
dissipation always spoil the exciton transfer from sites to the
reaction center, it is not difficult to understand why the
decoherence can not help the exciton transfer  within the
semi-classical approach.

\subsection{Quantum theory}
We now examine whether the quantum theory can explain the high
exciton transfer efficiency. For this purpose, we approximate
$\langle a_8^{\dagger} a_8 a_3^{\dagger} a_3\rangle$ as $\langle
a_8^{\dagger} a_8\rangle \langle a_3^{\dagger} a_3\rangle.$ Define
${n}_{mm}=\mbox{Tr}(\rho a_{m}^{\dagger} a_{m}), $ and
${n}_{mn}=\mbox{Tr}(\rho a_{m}^{\dagger} a_{n}).$ We obtain from Eq.
(\ref{masterE})
\begin{widetext}
\begin{eqnarray}
\dot{n}_{mm}&=&-i\sum_j{g_{mj}n_{mj}}+i\sum_j{g_{mj}n_{jm}}
-2\Gamma_m{n_{mm}}-\sum_j\Gamma_{mj}(n_{jm}+n_{mj}),\nonumber\\
\dot{n}_{mn}&=&i(\omega_m-\omega_n)n_{mn}+i\sum_j{g_{jm}}{n_{jn}}-i\sum_j{g_{jn}}{n_{mj}}
-(\Gamma_m+\Gamma_n+\gamma_m+\gamma_n-2\gamma_{mn})n_{mn}\nonumber\\
&&-\sum_j\Gamma_{mj}n_{jn}-\sum_j\Gamma_{nj}n_{mj}-a\Gamma_8{n_{mn}}(n_{88}+1),
m\neq{n},\nonumber\\
\dot{n}_{88}&=&2\Gamma_8{n_{33}}(n_{88}+1),
\label{fullQ}
\end{eqnarray}
\end{widetext}
where $a=1$ for $m=3$ or $n=3$, and $a=2$ for $m=n=3$, otherwise
$a=0$. Eq. (\ref{fullQ}) is accurate when $\langle a_8^{\dagger} a_8
a_3^{\dagger} a_3\rangle$ $=$ $\langle a_8^{\dagger} a_8\rangle
\langle a_3^{\dagger} a_3\rangle.$ In fact, our numerical
simulations show that this is exactly the case for small number of
exciton. For instance,  with $N_0=1$ or $2$, Monte Carlo simulations
show that $\langle a_8^{\dagger} a_8 a_3^{\dagger}
a_3\rangle-\langle a_8^{\dagger} a_8\rangle \langle a_3^{\dagger}
a_3\rangle=0.$  On the other hand, our model  backs to the two-level
model for the FMO complex when $N_0=1$\cite{caruso09}, and the
numerical  results given by  Eq. (\ref{fullQ}) is in agreement with
that  in \cite{caruso09}. It is difficult to prove $\langle
a_8^{\dagger} a_8 a_3^{\dagger} a_3\rangle-\langle a_8^{\dagger}
a_8\rangle \langle a_3^{\dagger} a_3\rangle=0$ for any $N_0$.
Fortunately, the site 3 and the reaction center 8 are connected only
through the irreversible process ${\cal L}_{8}(\rho)$, and this term
unlikely creates entanglement between site 3 and 8. Therefore if
site 3 and 8 are initially in a separable state, then they will
remain unentangled  forever. For site 3 and 8 in separable states,
$\langle a_8^{\dagger} a_8 a_3^{\dagger} a_3\rangle=\langle
a_8^{\dagger} a_8\rangle \langle a_3^{\dagger} a_3\rangle$ holds
true. This is not a proof, so we prefer to treat $\langle
a_8^{\dagger} a_8 a_3^{\dagger} a_3\rangle=\langle a_8^{\dagger}
a_8\rangle \langle a_3^{\dagger} a_3\rangle$ as an approximation.

\begin{figure}
\begin{minipage}[t]{0.9\linewidth}
\includegraphics*[width=1\columnwidth,
height=0.6\columnwidth]{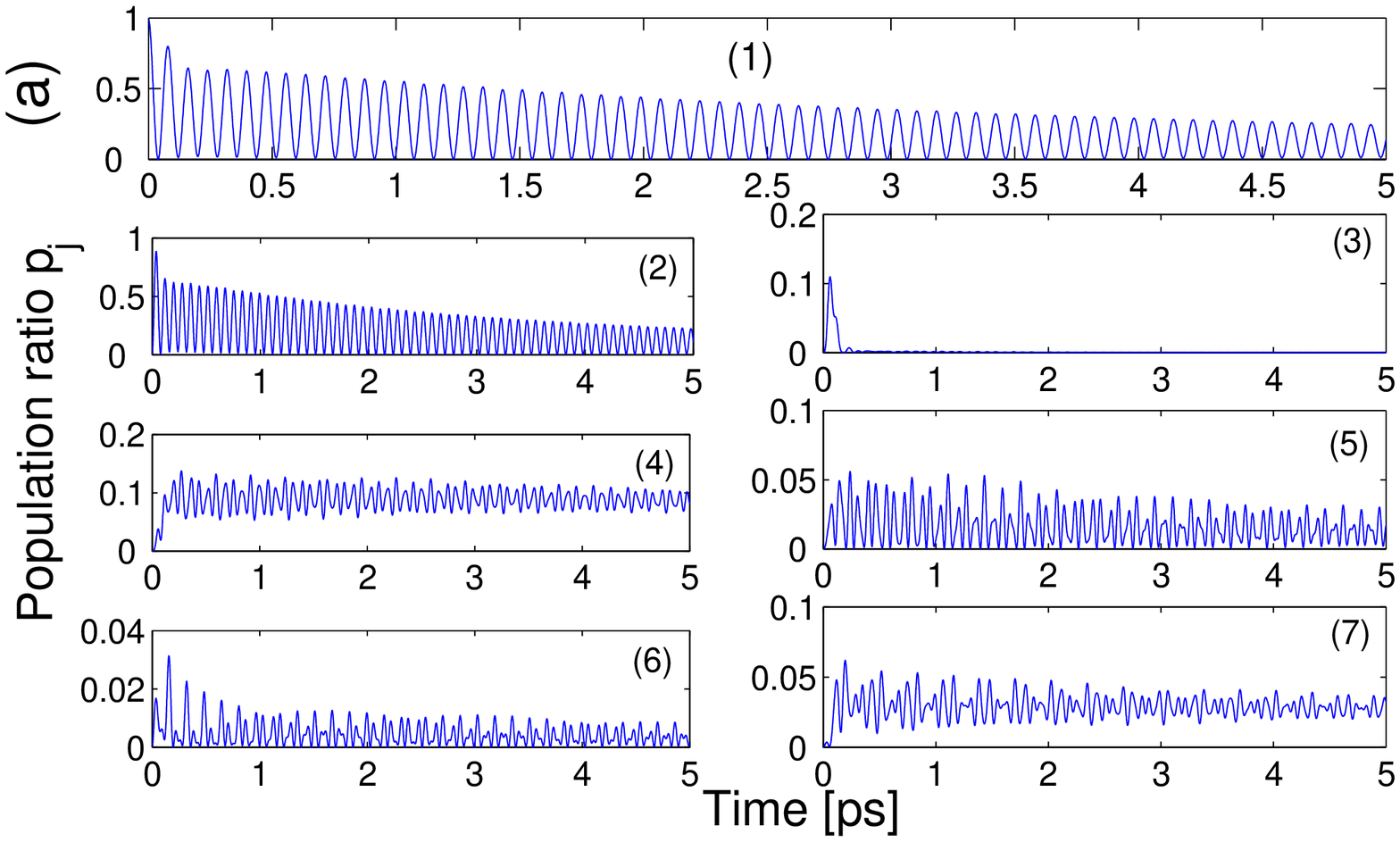}
\end{minipage}
\begin{minipage}[t]{0.9\linewidth}
\includegraphics*[width=1\columnwidth,
height=0.6\columnwidth]{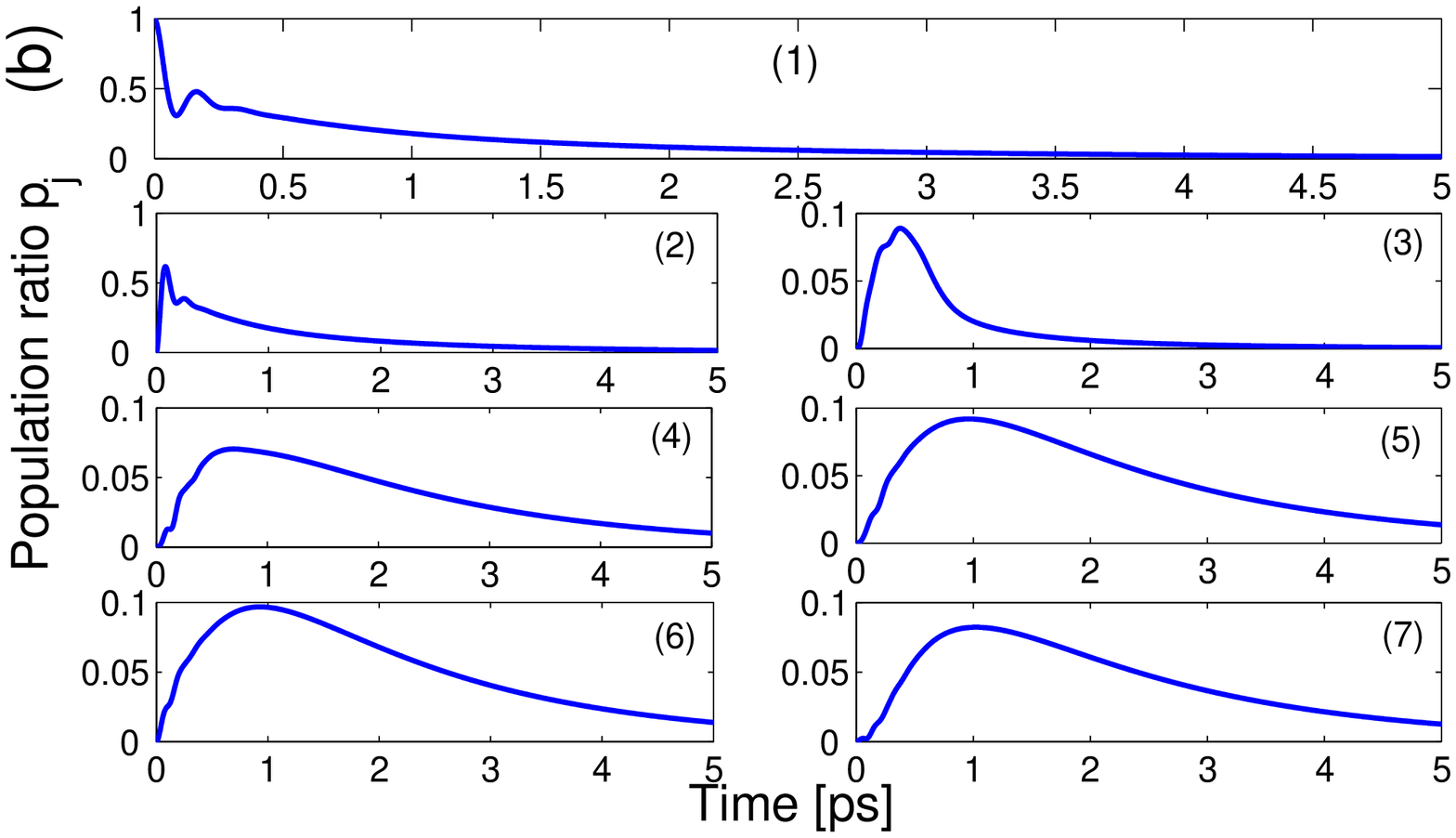}
\end{minipage}
\caption{Population ratio of $j$-th site defined by
$p_j=\frac{n_{jj}}{N_0}$ as  a function of time (number 1,2,...,7
label the sites). (a) is for the  semi-classical approach. And  (b)
is for the quantum theory. Dissipation and dephasing rates are the
same as those which optimize the efficiency with initial exciton
number $N_0=100$.} \label{fig4}
\end{figure}

We first study the case where only local decoherence exists, namely,
$\Gamma_{ij}=\gamma_{ij}=0.$  By numerically solving Eq.
(\ref{fullQ}), we find that without decoherence  the maximal energy
transfer efficiency is 67.06\% (see Fig. \ref{fig2}(b)), which is  a
bit larger than that given by the semi-classical approach. However,
with the assistance  of local decoherence, an efficiency over ninety
percent can be obtained (See Fig. \ref{fig3}(b)). In order to obtain
such a high energy transfer efficiency, we have to optimize the 15
decoherence rates ($\Gamma_j$, $\gamma_j$, and $\Gamma_8$,
$j=1,2,3,...,7)$, this is a time-consuming task. To save the
computer time,  we specify  the dissipation rates to
$\Gamma_j=0.0005$ for each site to  match  the measured life time in
experiments, which is the order of nanosecond \cite{adolphs06} for
the light-harvesting system. The optimal dephasing rates for highest
efficiency are found numerically as ($\gamma_1, \gamma_2, \gamma_3,
\gamma_4, \gamma_5, \gamma_6, \gamma_7$) =(0.74, 24, 0, 5.2, 50.6,
0, 15) and $\Gamma_8=0.32$, the corresponding efficiency is 91.77\%
for initial  $N_0=100$ on set 1.(see Fig. \ref{fig3}). Further
numerical simulations show  that the energy transfer efficiency can
reach over ninety percent for almost arbitrary exciton number $N_0$
(from 1 to 10000) on site 1, though the optimal dephasing rates are
different. With the optimal dephasing rates, the ratio of exciton
number on each site to the total number of exciton (defined as
$p_m=\frac{n_{mm}}{N_0}$) is plotted, see Fig.\ref{fig4} (b). We
find that after a rapid increasing (except for site 1), the exciton
number decreases  for each site, indicating  that most of the
exciton are transferred to the reaction center at site 8 (see Fig.
\ref{fig4}(b)). In contrast, the dynamics from  the semi-classical
approach (see Fig. \ref{fig4}(a)) is also plotted. More oscillations
can be observed, leading to less population transfer to the reaction
center.

\begin{figure}
\includegraphics*[width=0.75\columnwidth,
height=0.5\columnwidth]{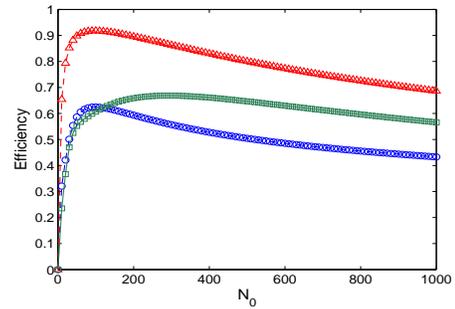} \caption{(color online)
Efficiency versus excitation number $N_0$ Parameters are chosen in
such a way that the efficiency is  optimized with $N_0=100$.
($\gamma_1, \gamma_2, \gamma_3, \gamma_4, \gamma_5, \gamma_6,
\gamma_7$) =(0.74, 24, 0, 5.2, 50.6, 0, 15), $\Gamma_8=0.32$ and
$\Gamma=0.0005$ is for the red triangle line. In contrast, we plot
the results without decoherence in green square and blue circle
lines  with $\Gamma_8=0.32$ (quantum) and $\Gamma_8=1.94$
(semi-classical), respectively. The later two $\Gamma_8$ optimize
the transfer efficiency for null decoherence case.}\label{fig5}
\end{figure}

It is interesting to note that the optimal decoherence rates
obtained for different exciton number $N_0$ are approximately  the
same values. To show this point we  have plotted in  Fig. \ref{fig5}
the efficiency as a function of the exciton number $N_0$, the
decoherence rates used in this figure are optimal for $N_0=100$.
Clearly, the decoherence rates optimal for $N_0=100$ can also result
in high transfer efficiency for a wide range of $N_0$. For instance,
the efficiency remains above ninety percent at $N_0=200$ with the
optimal decoherence rates for $N_0=100$. This observation suggests
that the high efficient energy transport, which has been found for a
fixed excitation number remains valid for other experimental and
natural operating conditions. Biologically, this means  the
light-harvesting system is robust against the number of photon
captured at site 1. Meanwhile, we observe that the dependence on
small $N_0$ (see Fig. \ref{fig5}) is stronger than the dependence on
large $N_0$, suggesting that the transfer efficiency is sensitive to
the variations of small $N_0$. This feature can be understood by
carefully examining Eq. (\ref{fullQ}). Clearly, the losing rate of
population on site 3 depends on the population in the reaction
center, $n_{88}$. As a back action, the gaining rate of population
on the reaction center depends on $n_{33}$.

\subsection{Effect of non-local decoherence}

\begin{figure}
\includegraphics*[width=0.8\columnwidth,
height=0.4\columnwidth]{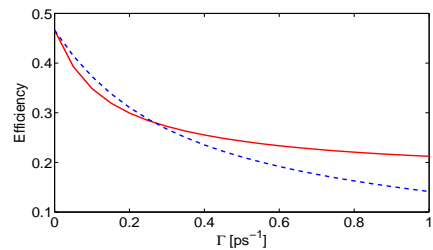} \caption{(color online)
Efficiency as a function of local dissipation rate
($\Gamma_j=\Gamma, j=1,2,...,7$). This figure is plotted to show the
effect of non-local decoherence on the transfer efficiency. Red
solid line represents a case with nonlocal dissipation
($\Gamma_{ij}=\Gamma$), whereas the  blue dash line is for the case
without nonlocal dissipation ($\Gamma_{ij}=0$), $i,j=1,2,...,7$.
Other parameters  chosen are $\Gamma_8=6$, $\gamma_j=0$, and
$\gamma_{ij}=0$.}\label{Ndis}
\end{figure}

In this section, we examine the effect of non-local decoherence on
the energy transfer efficiency and the dynamics of the  FMO complex.
When taking Eqs. (\ref{Ndiss}) and (\ref{Ndeph}) into account, we
must guarantee  that the non-local decoherence terms with rates
$\Gamma_{ij}$ and $\gamma_{ij}$  should  keep the positivity and
trace preserving of the density matrix \cite{caruso09}. For this
reason, we choose all $\Gamma_{ij}$ and $\gamma_{ij}$ positive to
optimize the transfer efficiency.  We find that the energy transfer
efficiency can be increased by  properly choosing   non-local
dephasing rates. For example, with  the optimized local decoherence
 rates (\{$\gamma_1, \gamma_2, \gamma_3, \gamma_4, \gamma_5,
\gamma_6, \gamma_7$\} =\{0.74, 24, 0, 5.2, 50.6, 0, 15\},
$\Gamma_8=0.32$, and $\Gamma=0.0005$), the efficiency can be
increased from 91.77$\%$ without non-local decoherence to 91.907$\%$
with non-local decoherence rates,
$(\gamma_{17},\gamma_{71},\gamma_{25},\gamma_{52})$=(0.74,0.74,24,24)
and $\gamma_{ij}=0, \{i,j\}\neq\{1,2,5,7\}$. Extensive numerical
simulations show  that energy transfer efficiency can not be
increased by weak non-local decoherence $\Gamma_{ij}$ (orders of
ns$^{-1}$), but strong non-local decoherence can improve the energy
transfer efficiency as  Fig. \ref{Ndis} shows.

\section{Conclusion}
In summary, we have studied the dynamics in light-harvesting
complexes beyond the single exciton limit and optimized the energy
transfer efficiency in the Fenna-Matthew-Olson (FMO) complex. To
describe the multi-excitation scenario, we have proposed a new model
for the propagation of excitation transfer, the model consists of 7
coupled cavities and a reaction center. Four types of decoherence,
including local dephasing, local dissipation, nonlocal dephasing and
nonlocal dissipation are considered. To match the life-time of
exciton observed in experiment, we fixed the local dissipation rates
to 0.0005 ps$^{-1}$. The local dephasing rates that optimizes the
transfer efficiency are given by numerical simulation. We find that
for multi-excitation case, the energy transfer efficiency can be
over 90\% under realistic conditions. Non-local decoherence can
slightly increase the efficiency, but it seems not important in the
light-harvesting mechanism. Moreover, we find that the transfer
efficiency is not sensitive to the initial excitation number at site
1. This suggests that the light-harvesting antenna may capture more
photons once and the experimental conditions are flexible to
simulate the light-harvesting in FMO complex.

\ \ \\
We are indebted to C. P. Sun for suggestions and stimulating talks.
Discussions with J. Cheng and S. Yi are acknowledged. This work is
supported by NSF of China under grant Nos 61078011 and 10935010, as
well as the National Research Foundation and Ministry of Education,
Singapore under academic research grant No. WBS: R-710-000-008-271.

\end{document}